\documentclass[english,runningheads]{llncs}

\usepackage{subcaption} 

\usepackage[latin1]{inputenc}
\usepackage[T1]{fontenc}
\usepackage{babel}
\usepackage{graphicx}
\usepackage{amsmath}
\usepackage{amsfonts}
\usepackage{amssymb}    %
\usepackage{multirow}
\usepackage{multicol}
\usepackage{breakcites}
\usepackage{url}
\usepackage{stmaryrd}
\usepackage[nomath]{lmodern}  
\usepackage{longtable} 
\usepackage{rotating}
\usepackage{array}
\usepackage{pdflscape}
\usepackage{soul} 
\usepackage{xspace}
\usepackage{enumitem}
\usepackage{hyperref}
\usepackage{color}
\usepackage{wrapfig}  
\usepackage{colortbl} 
\usepackage{fix-cm}  

\usepackage{proof}
\usepackage{makecell}

\usepackage{listings}

\newcommand{\TransfTo}{\Mapsto}   
\newcommand{\Cata}{${\mathcal T}_{\mathit{cata}}$}  
\newcommand{\mq}{${\mathcal T}_{\mathit{mq}}$}  

\newenvironment{sizepar}[2]
{\par\fontsize{#1}{#2}\selectfont}
{\par}

\newcommand{\us}{\raisebox{-3pt}{\hspace{-0.2mm}-\hspace{-0.5mm}-\hspace{-0.3mm}}} 
 
\newcommand{\If}{\leftarrow}

\newcommand{\plus}{\!+\!}

\newcommand{\false}{\mathit{false}}

\newcommand{\Down}{\raisebox{-.5mm}{\rule{0mm}{1mm}}}

\newcommand{\append}{\mathit{append}}
\newcommand{\snoc}{\mathit{snoc}}
\newcommand{\first}{\mathit{first}}
\newcommand{\last}{\mathit{last}}
\newcommand{\ordered}{\mathit{ordered}}
\newcommand{\inssort}{ins\us sort}
\newcommand{\ordins}{ord\us ins}
\newcommand{\emptylist}{empty\us list}
\newcommand{\nots}{\sim}
\newcommand{\ands}{\scalebox{0.8}[.9]{\&}}
\newcommand{\ors}{\scalebox{0.9}[.9]{$\vee$}}

\newcommand{\vericat}{{VeriCaT}\xspace}
\newcommand{\vericatmq}{{VeriCaT}$_{mq}$\xspace}

\title{Multiple Query Satisfiability \\of Constrained Horn Clauses\,\thanks{The 
Version of Record of this contribution 
will be published in the Proceedings of PADL 2023, Boston, MA, USA, January 
16--17, 2023: The 25th International Symposium on ``Practical Aspects of Declarative Languages'' as a volume of the Lecture Notes of Computer Science (LNCS) series of Springer Nature.}}

\titlerunning{Multiple Query Satisfiability of CHCs}

\author{
Emanuele De Angelis\inst{1}\orcidID{0000-0002-7319-8439}\and
Fabio~Fioravanti\inst{2}\orcidID{0000-0002-1268-7829} \and
Alberto~Pettorossi\inst{1,3}\orcidID{0000-0001-7858-4032}\and
Maurizio~Proietti\inst{1}\orcidID{0000-0003-3835-4931}
}

\authorrunning{E.~De~Angelis, F.~Fioravanti, A.~Pettorossi, and M.~Proietti}

\institute{IASI-CNR, Rome, Italy
	\email{\{emanuele.deangelis,maurizio.proietti\}@iasi.cnr.it}
	\and DEc,\,University\,`G.\,d'Annunzio',\,Chieti-Pescara,\,Italy
	\email{fabio.fioravanti@unich.it}
	\and DICII, University of Rome `Tor Vergata', Italy
	\email{pettorossi@info.uniroma2.it}
}

\begin{document}
\maketitle

\vspace*{-4mm}
\begin{abstract}
We address the problem of checking
the satisfiability of a set of constrained Horn clauses (CHCs)
possibly including more than one query.
We propose a transformation technique that 
takes as input a set of CHCs, including a set of queries,
and returns as output a new set of CHCs, such that
the transformed CHCs are satisfiable if and only if so are the original 
ones, and
the transformed CHCs incorporate in each new query 
suitable information coming from the other ones so that
the CHC satisfiability algorithm is able to exploit the relationships among all queries.
We show that our proposed technique is effective on a non trivial
benchmark of sets of 
CHCs that encode many verification problems for programs manipulating algebraic data types such as lists and trees.

\end{abstract}


\section{Introduction} 

\label{sec:Intro}
{\it Constrained Horn Clauses} (CHCs)
have been advocated as a logical formalism
very well suited for automatic program verification~\cite{Bj&15,DeAngelisFGHPP21}.
Indeed, many verification problems can be reduced to problems
of checking satisfiability
of CHCs~\cite{Gr&12}, and several effective CHC \textit{solvers}
are currently available as back-end tools for program verification
purposes~\cite{BlichaFHS22,De&14b,HoR18,Ko&16,KostyukovMF21}.

Following the CHC-based verification approach, a program is translated into
a set of definite CHCs
(that is, clauses whose head is different from~\textit{false}),
which capture the semantics of the program, together with
a set of queries (that is, clauses whose head is \textit{false}),
which specify the program properties to be verified.
Very often the CHC translation of the verification problem generates several queries.
In particular, this is the case when the program includes several functions,
each one having its own contract (that is,
a pair of a pre-condition and a
post-condition).

CHC solvers try to show the satisfiability of a set of CHCs of the form: $P \cup \{\false \!\leftarrow\! G_1,\ldots, \false \!\leftarrow\! G_n\}$,
where $P$ is a set of definite CHCs,
and $\false \!\leftarrow\! G_1,\ldots,\false \!\leftarrow\! G_n$ are queries,
by trying to show in a separate way the satisfiability of each set
$P\cup\{\false \!\leftarrow\! G_i\}$, for $i\!=\!1,\ldots,n$.
This approach may not be always quite effective, as
the solver may not be able to exploit the, possibly mutual, dependencies among the
various queries.
There is a simple way of combining all queries into one
(as done, for instance,
in the 
CHC solver competition~\cite{chccomp21}):
we introduce a new predicate $f$,
and from the above mentioned set of CHCs we get
$P\cup\{\false\!\leftarrow\!f, f\! \leftarrow\! G_1, \ldots,f\! \leftarrow\! G_n\}$.
However, also in this case, existing solvers will handle each query
separately and then combine the~results.

In this paper we propose, instead, a technique that, given a set
$P\cup\{\mbox{$\mathit{false} \!\leftarrow\! G_1,$}$ $\ldots,\false \!\leftarrow\! G_n\}$ of CHCs,
derives an {\it equisatisfiable} set
$P'\cup\{\false \!\leftarrow\! G'_1,\ldots,$ $\false \!\leftarrow\! G'_n\}$,
for whose satisfiability proof a CHC solver may exploit
the mutual interactions among 
the $n$ satisfiability proofs, one for each query.

Our 
technique builds upon the transformation approach
for verifying contracts that we presented in previous work and implemented in the
VeriCaT tool~\cite{DeAngelisFPP22}.
The algorithm of VeriCaT takes as input a
set of CHCs that manipulate algebraic data types (ADTs) such as lists
and trees,
and a set of CHCs defining contracts by means of
catamorphisms~\cite{MeijerFP91},
and returns as output a
set of CHCs without ADT variables such that the original set is satisfiable
if the new set is satisfiable.
For CHCs without ADT variables
state-of-the-art solvers are more effective in proving
satisfiability, and hence validity of contracts.
%

\indent
{The objective of the transformation algorithm
presented in this paper, which we call \mq,
is not to eliminate ADT variables, rather, it is to
incorporate into the clauses relative to a particular {query} 
some additional
constraints that are derived from {other queries}. 
These additional constraints are often very beneficial to the CHC
solvers when trying to check the satisfiability
of a given
set of clauses, thereby enhancing their ability to verify 
program properties.
Algorithm \mq\ is both sound and complete, that is,
the transformed clauses are satisfiable if and only if so are the original ones.
The completeness of \mq\ is very important because
if a property does \textit{not} hold, it allows
us to infer the unsatisfiability of the original clauses. Thus,
whenever the solver shows the unsatisfiability of the transformed clauses,
we deduce the invalidity of the property to be verified.

\vspace{-3mm}


\section{Preliminary Notions} 
\label{sec:CHCs}

We consider constrained Horn clauses that are defined in a
many-sorted first order language $\cal L$ with equality (=)
whose constraints are expressed using linear integer arithmetic (\textit{LIA})
and boolean (\textit{Bool\/}) expressions.
A {\em constraint} is a quantifier-free formula $c$, where
the \textit{LIA} constraints may occur as subexpressions of
boolean constraints, according to the SMT approach~\cite{Ba&09}:

\vspace{1.mm}
\hspace{-3mm}
$c~::=~d~~|~~\textit{B}~~|~~\textit{true}~~|~~\textit{false}~~|~\nots\!c~~|~~c_1 \,\ands\,c_2~~|~~c_1\!\ors c_2~~|~~c_1\!\Rightarrow\!c_2~|~~c_1\!=\!c_2~~|$~~


\hspace*{8mm}$\textit{ite}(c, c_1, c_2)~~|~~t\!=\!\textit{ite}(c, t_1, t_2)$
\vspace{1.mm}

\hspace{-3mm}
$d~::=~t_1\!\!=\!t_2~~|~~t_1\!\!<\!t_2~~|~~t_1\!\!\leq\!t_2~~|~~t_1\!\!\geq t_2~~|~~t_1\!\!>\!t_2 $

\vspace{1.mm}
\noindent
where:
(i)~{$B$} is a boolean variable, (ii)~$\nots,\ \ands$, \ors, and $\Rightarrow$
denote negation, conjunction, disjunction, and implication, respectively, (iii)~the ternary
function \textit{ite} denotes the if-then-else operator, and (iv)~{$t$} 
is a \textit{LIA}~term of the form {$a_0+a_1X_1+\dots+a_nX_n$}
with integer coefficients {$a_0,\dots,a_n$}
and variables {$X_1,...,X_n$}.
The equality symbol will be used both for integers and booleans.
We will often write $B\!=\!\mathit{true}$ 
(or $B\!=\!\mathit{false}$) as $B$ (or $\nots\!\!B$).
The theory of \textit{LIA} and boolean constraints will be denoted by
$\mathit{LIA\cup Bool}$.
The integer and boolean sorts are said to be 
{\it basic sorts}. 
A recursively defined sort (such as the sort of lists and trees) 
is said to be an {\it algebraic data type} (ADT, for short).

An {\it atom} is a formula of the form $p(t_{1},\ldots,t_{m})$, where~$p$
is a predicate symbol not occurring in $\textit{LIA}\cup\textit{Bool\/}$,
and $t_{1},\ldots,t_{m}$ are first order terms in $\cal L$.
A~{\it constrained Horn clause}  (CHC), or simply, a {\it clause}, is
an implication of the form $H\leftarrow c, G$.
The conclusion $H$, called the {\it head\/}, is either an atom or \textit{false}, and
the premise, called the {\it body\/}, is the conjunction of a constraint~$c$
and a conjunction~$G$ of zero or more atoms.
A clause is said to be a {\it  query\/} if its head is \textit{false}, and
a {\it definite clause\/}, otherwise.
Without loss of generality, we assume that every atom of the body
of a clause has distinct variables (of
any sort) as arguments. 
%
The set of all variables occurring in an {expression} $e$ is denoted by ${\it vars}(e)$.
By $\mathit{bvars(e)}$ (or $\mathit{adt}$-$\mathit{vars}(e)$) we
denote the set of variables in~$e$ whose sort is a basic sort (or an ADT sort).
The {\it universal closure} of a formula~$\varphi$ is denoted by $\forall (\varphi)$.

Let~$\mathbb D$ be the usual interpretation for the symbols of {theory}
$\textit{LIA}\cup\textit{Bool\/}$.
By $M(P)$ we denote the {\it least} ${\mathbb D}$-model of a set $P$ of definite  clauses~\cite{JaM94}.
%


Now, in order to characterize the class of queries that can be handled using our transformation
technique, we introduce (see Definition~\ref{fig:cataExt} below) a class of recursive schemata defined by CHCs~\cite{DeAngelisFPP22}.
That class 
is related to those of \textit{morphisms}, \textit{catamorphisms}, and
\textit{paramorphisms} considered in functional programming~\cite{HinzeWG13,MeijerFP91}.
We will not introduce a new terminology here and we will refer to our schemata as
\textit{generalized catamorphisms},
or {\it catamorphisms}, for short.

Let $f$ be a predicate symbol with $m \plus n$ arguments (for $m\!\geq\! 0$ and $n\!\geq\! 0$) whose sorts are
\! $\alpha_1,\!\ldots\!,\!$ $\alpha_m,$ $\beta_1,\ldots,\beta_n$, respectively.
We say that $f$ is a {\em functional predicate} from $\alpha_1\times\ldots\times\alpha_m$ to $\beta_1\times\ldots\times\beta_n$,
{with respect to a set $P$ of definite clauses},
if $M(P)\! \models\! \forall X,\!Y,\! Z.\ f(X,\!Y) \wedge f(X,\!Z) \rightarrow (Y\!=\!Z)$,
where $X$ is an $m$-tuple of distinct variables, and
$Y$ and $Z$ are $n$-tuples of distinct variables.
Given the atom $f(X,Y)$, we say that
$X$ and $Y$ are the tuples of the {\em input}
and {\em output} variables of $f$, respectively.
Predicate~$f$ is said to be {\em total} 
if $M(P) \models \forall X \exists Y.\ f(X,Y)$.
In what follows, a `total, functional predicate' $f$ from a tuple~$\alpha$ of sorts to a tuple~$\beta$ of sorts
will be called a `total function' 
and denoted by $f \in \mbox{[}\alpha \rightarrow \beta\mbox{]}$
(the set $P$ of clauses that define $f$ will be understood from the context).

\vspace*{-1mm}
\begin{definition}[Generalized Catamorphisms]\label{def:cata}
A {\em generalized list catamorphism}, shown in Figure~{\rm\ref{fig:cataExt}~(A)}, is a
total function $h\in [\sigma\times\mathit{list}(\beta)
\rightarrow \varrho]$, where\,$:$ $(i)~\sigma$, $\beta$,
$\varrho$, and $\tau$ are $($products of$\,)$ basic sorts,
$(ii)~\mathit{list}(\beta)$ is the sort of lists of elements of sort $\beta$,
$(iii)~\mathit{base}1$ is a total function in $[\sigma\rightarrow \varrho]$,
$(iv)~f$ is a catamorphism in $[\sigma\times\mathit{list}(\beta)
\rightarrow \tau]$
and $(v)$~$\mathit{combine}1$ is a total function in
$[\sigma\times\beta\times\varrho\times\tau\rightarrow \varrho]$.
Similarly, a {\em generalized tree catamorphism}
is a total function
$t\in [\sigma\times\mathit{tree}(\beta)\rightarrow \varrho]$ defined
as shown in Figure~{\rm\ref{fig:cataExt}~(B)}.
\begin{figure}[ht!]
\vspace*{-1mm}
\begin{sizepar}{9}{11}
\begin{center}
\hspace*{-4mm}
(A)~ \begin{subfigure}{.35\textwidth}
{
$h(X,[\,],Y) \If   \mathit{base}1(X,Y).\\
h(X,[H|T],Y) \If \\
\hspace*{5mm}f(X,T,\mathit{Rf}),\\
\hspace*{5mm}h(X,T,R),\\
\hspace*{5mm}\mathit{combine}1(X,H,R,\mathit{Rf},Y).$
}
\end{subfigure}\hspace*{-5mm}
\hspace*{5mm}
(B)~ \begin{subfigure}{.45\textwidth}
{
$t(X,\mathit{leaf},Y) \If  \mathit{base}2(X,Y).\\
t(X,\mathit{node}(L,N,R),Y) \If  \\
\hspace*{5mm} g(X,L,\mathit{RLg}), \ \ g(X,R,\mathit{RRg}),\\
\hspace*{5mm} t(X,L,\mathit{RL}), \ \ t(X,R,\mathit{RR}), \\
\hspace*{5mm} \mathit{combine}2(X,N,\mathit{RL,RR,RLg,RRg},Y).$
}
\end{subfigure}
\end{center}
\end{sizepar}
\vspace{-7mm}
\caption{\rm (A) Generalized list catamorphism. \,(B) Generalized tree catamorphism.\label{fig:cataExt}}
\vspace*{-4mm}
\end{figure}
\end{definition}
Note that the above definition is recursive,
that is, the predicates $f$ and $g$ are defined by instances of schemata (A) and (B), respectively.
Examples of catamorphisms will be shown in the following section.

\section{An Introductory Example} 
\label{sec:IntroExample}
Let us consider the clauses of Figure~\ref{fig:chc_prog}, which are the result of
translating an iterative program for Insertion Sort. (Details
on how this translation can be performed are outside the scope of this paper.)
The clauses in Figure~\ref{fig:chc_prog} will be called {\em program clauses}
and the predicates defined by those clauses will be called {\em program predicates}.


We have that $\mathit{\inssort(Xs,Ys,Zs)}$ 
holds if $\mathit{Zs}$ is
the ordered list of integers (here and in what follows, the order is with respect to~$\leq$) obtained by inserting
every element of the list $\mathit{Ys}$ in the proper position of
the ordered list $\mathit{Xs}$.
Thus, the result of sorting a list $\mathit{Ys}$
is the list $\mathit{Zs}$  such that $\mathit{\inssort([~],Ys,Zs)}$ holds.
The predicate $\mathit{\inssort}$ depends (see clause~2)
on the predicates $\mathit{\emptylist}$ and
$\mathit{\ordins}$.
We have that $\mathit{\emptylist(L)}$ holds if the list $\mathit{L}$ is empty,
and \linebreak $\mathit{\ordins(Y,Xs{\rm1},Xs{\rm2},Ys,Zs)}$ holds if:
(i)~the concatenation of the lists $\mathit{Xs{\rm1}}$ and $\mathit{Xs{\rm2}}$
is ordered,
(ii)~$Y$ is greater than or equal to the last element of
$\mathit{Xs{\rm1}}$, and
(iii)~$\mathit{\inssort(Xs',Ys,Zs)}$ holds,
where $\mathit{Xs'}$ is the concatenation of $\mathit{Xs{\rm1}}$ and
the ordered list obtained by inserting (according to the $\leq$ order)
$Y$ into~$\mathit{Xs{\rm2}}$.
As usual, $\mathit{append(Xs,Ys,Zs)}$ holds if $\mathit{Zs}$ is the
concatenation of the lists $\mathit{Xs}$ and $\mathit{Ys}$, and
$\mathit{snoc(Xs,Y,Zs)}$ holds if
$\mathit{append(Xs,[Y],Zs)}$ holds.

It is not immediate to see why
the above properties for $\mathit{\inssort}$ and $\mathit{\ordins}$
are valid. This is 
also due to the fact that 
the predicates $\mathit{\inssort}$ and $\mathit{\ordins}$
are mutually recursive.
The goal of this paper is to present a technique based on
CHC transformations that allows us to automatically
prove properties expressed by possibly mutually recursive predicates, using a CHC solver. 

\begin{figure}[ht!]
\vspace*{-9mm}
 \begin{equation*}
  \begin{split}
   \begin{aligned}
  &1.~\mathit{\inssort(Xs,[~],Xs).}        \\[-2pt]
  &2.~\mathit{\inssort(Xs,[Y|Ys],S) \leftarrow} \\[-2pt]
  &\hspace{10mm}\mathit{\emptylist(L),} \\[-2pt]
  &\hspace{10mm}\mathit{\ordins(Y,L,Xs,Ys,S).} \\[ -1pt]
  &3.~\mathit{append([~],Ys,Ys).}   \\[-2pt]
  &4.~\mathit{append([X|Xs],Ys,[X|Zs]) \leftarrow} \\[-2pt]
  &\hspace{10mm}\mathit{append(Xs,Ys,Zs).} \\[-1pt]
  &5.~\mathit{snoc([~],X,[X]).}     \\[-2pt]
  &6.~\mathit{snoc([H|T],X,[H|TX]) \leftarrow}     \\[-2pt]
  &\hspace{10mm}\mathit{snoc(T,X,TX).}     \\[ -1pt]
  &7.~ \mathit{\emptylist([~]).}   \\[-2pt]
   \end{aligned}
  \end{split}
  \qquad\qquad
  \begin{split}
   \begin{aligned}
  &~8.~\mathit{\ordins(Y,Xs{\rm 1},[~],Ys,Zs)}\leftarrow      \\[-2pt]
  &\hspace{7mm}\mathit{snoc(Xs{\rm1},Y,Xs{\rm1}\!Y),}      \\[-2pt]
  &\hspace{7mm}\mathit{\inssort(Xs{\rm1}\!Y\!,Ys,Zs).}  \\[-2pt]
  &~9.~\mathit{\ordins(Y,Xs{\rm1},[X|Xs{\rm2}],Ys,Zs)\leftarrow} \\[-2pt]
  &\hspace{7mm}\mathit{Y\!\leq\! X,} \ \mathit{snoc(Xs{\rm1},Y,Xs{\rm1}\!Y),}    \\[-2pt]
  &\hspace{7mm}\mathit{snoc(Xs{\rm1}\!Y,X,Xs{\rm1}\!YX),}   \\[-2pt]
  &\hspace{7mm}\mathit{append(Xs{\rm1}\!YX,Xs{\rm2},Xs),} \\[-2pt]
  &\hspace{7mm}\mathit{\inssort(Xs,Ys,Zs).}  \\[-2pt]
  &10.~\mathit{\ordins(Y,Xs{\rm1},[X|Xs{\rm2}],Ys,Zs)}\leftarrow \\[-2pt]
  &\hspace{7mm}\mathit{Y\!>\!X,} \  \mathit{snoc(Xs{\rm1},X,Xs{\rm1}X),}  \\[-2pt]
  &\hspace{7mm}\mathit{\ordins(Y,Xs{\rm1}X,Xs{\rm2},Ys,Zs).}  \\[3pt]
   \end{aligned}
  \end{split}
 \end{equation*} \vspace*{-4mm}
 \caption{Program clauses for Insertion Sort.
 \label{fig:chc_prog}}
 \vspace{-7mm}
\end{figure}

Now we will present the clauses that formalize the properties 
we want to show in our Insertion Sort example.
This set of clauses
is made out of two subsets: $(A)$~a set of definite CHCs that define the
catamorphisms used
in the queries, and $(B)$~a set of queries that specify the program properties
to be shown.

For~$(A)$, in Figure~\ref{fig:chc_catas}
we present the definite CHCs defining the three catamorphisms we will use. They are:
(i)~$\mathit{ordered}$, which takes as input a list~$\mathit{Ls}$ and returns a boolean
$\mathit{B}$ such that if $\mathit{Ls}$ is ordered, 
then $\mathit{B}\!=\!\mathit{true}$, otherwise $\mathit{B}\!=\!\mathit{false}$;
(ii)~$\mathit{first}$, which takes as input a list $\mathit{Ls}$ and returns
 a boolean~$\mathit{B}$ and an element $F$ such that
if $\mathit{Ls}$ is empty, then $\mathit{B}\!=\!\mathit{false}$ and
$\mathit{F}\!=\!0$ (this value for~$F$ is an arbitrary integer and will not be used elsewhere),
otherwise $\mathit{B}\!=\!\mathit{true}$ and
$\mathit{F}$ is the head of $\mathit{Ls}$; and
(iii)~$\mathit{last}$, which is analogous to $\mathit{first}$, except that it returns
the last element~$\mathit{L}$, if any, instead of the first element.

The predicates of these catamorphisms are called {\it property predicates}.


\begin{figure}
\vspace*{-9mm}
 \begin{equation*}
  \begin{split}
   \begin{aligned}
  &11.~\mathit{ordered([~],B) \leftarrow B.}       \\[-2pt]
  &12.~\mathit{ordered([H|T],B) \leftarrow}        \\[-2pt]
  &\hspace{7mm}\mathit{B= (B{\rm1} \Rightarrow (H\!\leq\!F ~\ands~ B{\rm2})),} \\[-2pt]
  &\hspace{7mm}\mathit{first(T,B{\rm1},F), ordered(T,B{\rm2}).}    \\[ 1pt]
  &\\
   \end{aligned}
  \end{split}
  \quad \quad
  \begin{split}
   \begin{aligned}
  &13.~\mathit{first([~],B,F)    \leftarrow  ~\sim\!\!B ~\ands~ F\!=\!{\rm 0}.  } \\[-2pt]
  &14.~\mathit{first([H|T],B,F) \leftarrow  B ~\ands~ F\!=\!H.}    \\[2pt]
  &15.~\mathit{last([~],B,L)    \leftarrow  ~\sim\!\!B ~\ands~ L\!=\!{\rm 0}.   } \\[-2pt]
  &16.~\mathit{last([H|T],B,L) \leftarrow  }    \\[-3pt]
  &\hspace{7mm}\mathit{B ~\ands~ L\!=\!ite(B{\rm1},L{\rm1},H), last(T,B{\rm1},L{\rm1}).      }
   \end{aligned}
  \end{split}
 \end{equation*} \vspace*{-5mm}
 \caption{Property clauses for Insertion Sort: (A) the catamorphisms.
 \label{fig:chc_catas}}
 \vspace*{-6mm}
\end{figure}


For $(B)$, we present the set of queries that specify the
program properties.
We assume that: (i)~at most one property is specified
for each program predicate,
and (ii)~the property related to the program predicate $p$ is expressed as an implication of the form:
$p(\ldots),~  \mathit{cata}_{1}(\ldots),~ \ldots,~ \mathit{cata}_{n}(\ldots)
\rightarrow d$,
where 
the $\mathit{cata}_{i}(\ldots)$'s are catamorphisms and $d$ is a constraint.
This implication can be expressed as a query
by adding $\sim\!d$ to its premise and changing
its conclusion to $\mathit{false}$.
For our Insertion Sort example, in  Figure~\ref{fig:chc_queries}
we specify four properties by introducing a query for the predicates
$\inssort$, $\ordins$, $\mathit{append}$, and $\mathit{snoc}$.

For $\inssort$, query $\mathit{q}1$ states that given a list of integers $\mathit{Ys}$,
if $\mathit{Xs}$ is ordered 
and  $\mathit{\inssort(Xs,\!Ys,Zs)}$ holds,
then $\mathit{Zs}$ is ordered.
For $\ordins$, query $\mathit{q}2$ states that if the concatenation of the lists
$\mathit{Xs{\rm1}}$ and $\mathit{Xs{\rm2}}$ is ordered (that is, $\mathit{Xs{\rm1}}$
and $\mathit{Xs{\rm2}}$ are ordered, and the last element of $\mathit{Xs{\rm1}}$
is less than or equal to the first element of $\mathit{Xs{\rm2}}$),
$\mathit{Y}$ is greater than or equal to the last element of $\mathit{Xs{\rm1}}$,
and $\mathit{\ordins(Y,Xs{\rm1},Xs{\rm2},Ys,Zs)}$ holds,
then $\mathit{Zs}$ is ordered.
For $\mathit{append}$, query $\mathit{q}3$ states that if $\mathit{Xs{\rm1}}$ and $\mathit{Xs{\rm2}}$
are ordered, and the last element of $\mathit{Xs{\rm1}}$ is less than or
equal to the first element of $\mathit{Xs{\rm2}}$
and $\mathit{append(Xs,\!\!Ys,\!Zs)}$ holds, then $\mathit{Zs}$ is ordered.
For $\mathit{snoc}$, query $\mathit{q}4$ states that if $\mathit{Xs}$ is ordered, and the
last element of $\mathit{Xs}$ is less than or equal to $\mathit{X}$,
and $\mathit{snoc(Xs,X,XsX)}$ holds, then $\mathit{XsX}$ is ordered.


\begin{figure}
\vspace*{-5mm}
 \begin{equation*}
  \begin{aligned}
(q1)~~~ & \mathit{false \leftarrow ~ \sim\!(B{\rm1} \Rightarrow B{\rm2}),\
   ordered(Xs,B{\rm1}),\ ordered(Zs,B{\rm2}),
   \inssort(Xs,Ys,Zs).}       \\[-1pt]
   (q2)~~~ & \mathit{false \leftarrow ~ \sim\!((B{\rm1} ~\ands~ B{\rm2}
   ~\ands~ (( B{\rm4} ~\ands~ B{\rm5} ) \Rightarrow L\leq F) ~\ands~
   ( B{\rm4} \Rightarrow L\leq Y )) \Rightarrow B{\rm3}),} \\[-2pt]
 & \mathit{~~ordered(Xs{\rm1},B{\rm1}),\ ordered(Xs{\rm2},B{\rm2}),\
 ordered(Zs,B{\rm3}),}       \\[-2pt]
 & \mathit{~~ last(Xs{\rm1},B{\rm4},L),\ first(Xs{\rm2},B{\rm5},F),\
 \ordins(Y,Xs{\rm1},Xs{\rm2},Ys,Zs).}    \\[-1pt]
   (q3)~~~ & \mathit{false \leftarrow ~ \sim\!((B{\rm1} ~\ands~ B{\rm2}
   ~\ands~ ((B{\rm4} ~\ands~ B{\rm5}) \Rightarrow L\leq F))
   \Rightarrow B{\rm3}),}   \\[-2pt]
 & \mathit{~~ordered(Xs,B{\rm1}),\ ordered(Ys,B{\rm2}),\ ordered(Zs,B{\rm3}),} \\[-2pt]
 & \mathit{~~ last(Xs,B{\rm4},L),\ first(Ys,B{\rm5},F),\ append(Xs,Ys,Zs).}    \\[-1pt]
   (q4)~~~ & \mathit{false} \leftarrow ~ \sim\!((B{\rm1} ~\ands~ (B{\rm2}
    \Rightarrow L\leq X)) \Rightarrow B{\rm3}),\     \\[-2pt]
 & \mathit{~~ordered(Xs,B{\rm1}),\ last(Xs,B{\rm2},L),\ ordered(XsX,B{\rm3}),
  snoc(Xs,X,XsX).}
  \end{aligned}
 \end{equation*}
 \vspace*{-4mm}
 \caption{Property clauses for Insertion Sort: (B) the queries.  \label{fig:chc_queries}}
 \vspace*{-6mm}
\end{figure}

\smallskip
At this point, we can check whether or not the properties expressed
by the queries \mbox{$q1$--$q4$} do hold
by checking the satisfiability of the program clauses together with the
property clauses.
In order to perform that satisfiability check, we have used
the state-of-the-art CHC solver SPACER, based on Z3~\cite{Ko&16}.
SPACER failed to return an answer within five minutes.
The weakness of CHC solvers for examples  like the one presented here,
motivates our technique based on CHC transformation.
This technique produces an equisatisfiable set of CHCs whose
satisfiability can be, hopefully, easier
to verify.
Indeed, in our example, SPACER
succeeds to prove the satisfiability of
the new set of CHCs produced by our transformation algorithm.
In Section~\ref{sec:Experiments}, we will show that our technique
improves the effectiveness of state-of-the-art solvers on a non-trivial benchmark.

%
%


\section{Catamorphism-based Queries} 
\label{sec:Cata}

\vspace{-1mm}
As already mentioned, the translation of a program verification problem to CHCs usually  
generates two disjoint sets of clauses: (i)~the set of {\it program 
clauses}, and
(ii)~the set of {\it property clauses}, with the associated sets of
 {\em program predicates}, and 
 {\em property predicates}. 
Without loss of generality, we will assume 
that property predicates may occur in the property clauses only.
Moreover, in order to define a class of CHCs where our transformation technique
always terminates, we will consider property predicates that are {\it catamorphisms}
(see Definition~\ref{def:cata}). 
In the sequel we need the following definitions. 
An atom is said to be a {\textit{program atom}} (or a {\textit{catamorphism atom}}), 
if its predicate symbol is a program predicate (or a catamorphism, respectively).
Recall 
that when writing a catamorphism atom as $\mathit{cata}(X,T,Y)$, we stipulate that
$X$ is the (tuple of its) input basic variable(s), $T$ is the input ADT variable, and $Y$ is
the (tuple of its) output basic variable(s).


\vspace{-1mm}
\begin{definition}\label{def:query}
A \emph{catamorphism-based query}  
is a query of the form\,$:$\nopagebreak

\hspace{8mm}$\mathit{false} \ \leftarrow c,\
\mathit{cata}_1(X_1,T_1,Y_1),\ \ldots,\ \mathit{cata}_n(X_n,T_n,Y_n),\  
\mathit{pred}(Z)$ 

\noindent
where$:$ $(i)$~$\mathit{pred}$ is a program predicate and $\mathit{Z}$ is a tuple of distinct variables,
$(ii)$\,$c$~is a constraint such that $\mathit{vars}(c)\!\subseteq\! 
\{X_{1},\ldots,X_{n},{Y_1},\ldots,{Y_n},Z\}$, 
$(iii)$~$\mathit{cata}_1,$ $\ldots,$ $\mathit{cata}_n$ are catamorphism
atoms,
$(iv)$~{${Y_1},\ldots,{Y_n}$ are pairwise} disjoint tuples of distinct variables of basic sort
not occurring in $X_{1},\ldots,X_{n}, Z$,
$(v)$~${T_1},\ldots,{T_n}$ are ADT variables 
occurring in~$Z$. 
\end{definition}
\vspace{-1mm}

The queries of Figure~\ref{fig:chc_queries} 
are examples of catamorphism-based queries.
Many interesting program properties can be defined as catamorphism-based queries, 
although, in general, this might require some ingenuity.
\vspace{-2mm}


\section{Transformation Rules} 
\label{sec:Rules}

\vspace*{-1mm}
In this section we present the rules that we use for transforming CHCs. 
These rules are variants of the usual transformation rules for 
CHCs (and CLP programs), specialized to our context where we use
catamorphisms. Then,
we prove the soundness and completeness of those rules.

The goal of the transformation rules is to incorporate catamorphisms into program predicates,
that is, to derive  for each program predicate $p$, a new predicate \textit{newp} 
{whose definition is given by the conjunction of an atom for $p$ with the catamorphism atoms}
needed for showing the satisfiability of the 
query relative to $p$. 
In this section we will 
indicate how this can be done referring 
our Insertion Sort example, while in the next section we will
present a transformation algorithm to perform this task in an automatic way.

A {\it transformation sequence from} $S_{0}$ {\it to} $S_{n}$
is a sequence 
$S_0 \TransfTo S_1 \TransfTo \ldots \TransfTo S_n$ of sets of CHCs 
such that, for $i\!=\!0,\ldots,n\!-\!1,$ $S_{i+1}$ is derived from $S_i$, 
denoted
$S_{i} \TransfTo S_{i+1}$, by performing a transformation step consisting in
applying one of the following \mbox{rules~R1--R4}.

\medskip
\noindent
(R1)~{\it Definition Rule.}
Let $D$ be a clause of the form
$\mathit{newp}(X_1,\ldots,X_k)\leftarrow c,\textit{Catas}, A$, 
where:
(1)~\textit{newp} is a predicate symbol not occurring in 
the sequence $S_0\TransfTo S_1\TransfTo$ $\ldots\TransfTo S_i$ constructed so far, (2)~$\{X_1,\ldots,X_k\} = \mathit{vars}(\{\textit{Catas},A\})$, 
(3)~$c$ is a constraint such that 
$\mathit{vars}(c)\! \subseteq\! \mathit{vars}(\{\textit{Catas},A\})$, 
(4)~\textit{Catas} is a conjunction of catamorphism atoms, with 
$\textit{adt-vars}(\textit{Catas})\subseteq \textit{adt-vars}(A)$, and
(5)~$A$ is a program atom.
By {\it definition introduction} we may add $D$ to $S_i$ and get $S_{i+1}= S_i\cup \{D\}$. 

\smallskip
The case where $A$ is absent is accommodated by considering 
$A$ to be 
$true(X)$, which holds for every $X$ of ADT sort.

For $j\!=\!0,\ldots, n$, by $\mathit{Defs}_j$ 
we denote the set of clauses, 
called {\it definitions}, 
introduced by rule~R1 during the construction of the
sequence $S_0\TransfTo 
S_1\TransfTo \ldots\TransfTo S_j$. 
Thus, $\mathit{Defs}_0\!=\!\emptyset$, and for $j\!=\!0,\ldots, n$, 
$\mathit{Defs}_j\!\subseteq\!\mathit{Defs}_{j+1}$.

\smallskip
In our Insertion Sort example, the set $S_0$ consists of all the clauses 
shown in Figures~\ref{fig:chc_prog}, \ref{fig:chc_catas}, and \ref{fig:chc_queries}, and we start off by introducing
 the following definition
(with constraint \textit{true}), whose body consists of the atoms in the body of query $q1$: 

\vspace{.5mm}
\noindent $D1$.~$\mathit{new{\rm1}(\!X\!s,\!B{\rm1},\!Z\!s,\!B{\rm2},\!\!Y\!\!s)\!
\leftarrow\! \ordered(\!X\!s,\!B{\rm1}\!),  
\ordered(\!Z\!s,\!B{\rm2}), \inssort(\!X\!s,\!\!Y\!\!s,\!Z\!s\!).}$

\vspace{.5mm}
\noindent
Thus, $S_1 = S_0 \cup \{D1\}$.~\hfill$\Box$~~

\smallskip

The clauses for \textit{newp} are obtained by first
(i)~\textit{unfolding} the definition of \textit{newp}, then 
(ii)~incorporating some catamorphisms and constraints provided by the queries 
into the clauses derived by unfolding, 
and (iii)~finally, folding using suitable new definitions.
Now, we introduce an unfolding rule (see R2 below), which actually is
the composition of the unfolding and the application of the functionality
property presented in previous work~\cite{DeAngelisFPP22}.
Let us first define the notion of the one-step unfolding which is a step of 
symbolic evaluation performed by applying once the resolution rule.
\vspace{-1mm}
\begin{definition}[\mbox{\it{One-step Unfolding}}]
Let  $C$: $H\leftarrow c,L,A,R$ be a clause, where $A$ is an atom,
and let $P$ be a set of definite clauses with 
$\mathit{vars}(C)\cap\mathit{vars}(P)=\emptyset$.
Let {\it Cls}: $\{K_{1}\leftarrow c_{1},
B_{1},~\ldots,~K_{m}\leftarrow c_{m}, B_{m}\}$, with $m\!\geq\!0$,
be the set of clauses in $P$,
such that: for $j=1,\ldots,m$,
$(i)$~there exists a most general unifier~$\vartheta_j$ of $A$ 
and~$K_j$, and {$(ii)$~the conjunction of constraints $(c, c_{j})\vartheta_j$ is satisfiable.}
The {\em{one-step unfolding}} produces the following set of CHCs\,$:$

$\mathit{Unf}(C,A,P)=\{(H\leftarrow  c, {c}_j,L, B_j, R) 
\vartheta_j \mid  j=1, \ldots, m\}$. 
\end{definition}
\vspace{-1mm}

In the following Rule R2 and in the sequel, 
{\it{Catas}} denotes a conjunction of catamorphism atoms.

\smallskip
\noindent
(R2)~{\it Unfolding Rule.} 
Let $D$: $\mathit{newp}(U) \If c, \mathit{Catas}, A$
be a definition in $S_i \cap \textit{Defs}_i$ and $P$ be the set of definite clauses in $S_i$.
We derive a new set \textit{UnfCls} of clauses by the following three steps.

\noindent
{\it Step}~1.\hspace{1mm}(\!{\it{One-step unfolding of the program atom}}) 
$\mathit{UnfCls}:=\mathit{Unf}(D,A,P)$;

\noindent
{\it Step}~2.\hspace{1mm}(\!{\it{Unfolding of the catamorphism atoms}})

\hspace{2mm}\begin{minipage}[t]{114mm}
\noindent 
{\bf while} there exists a clause $E$: 
\mbox{$H\! \leftarrow d, {L}, C, {R}$} in $\mathit{UnfCls}$, for some conjunctions~$L$ and~$R$ of atoms,
such that $C$ is a catamorphism atom whose argument of ADT sort is not a variable {\bf do}\\
\hspace*{4mm}$\mathit{UnfCls}:=(\mathit{UnfCls}\setminus\{E\}) \cup \mathit{Unf(E,C,P)}$;~
\end{minipage}

\noindent
{\it Step}~3.\hspace{1mm}(\!{\it{Applying Functionality}}) 

\hspace{2mm}\begin{minipage}[t]{114mm}
\noindent
{\bf while} there exists a clause $E$: \mbox{$H \leftarrow d, {L}, \mathit{cata}(X,T,Y1), \mathit{cata}(X,T,Y2),{R}$} 
 in $\mathit{UnfCls}$,  for some catamorphism $\mathit{cata}$ {\bf do}\\ 
\hspace*{4mm}$\mathit{UnfCls}:=(\mathit{UnfCls}-\{E\}) \cup \{H \leftarrow d, Y1\!=\!Y2, {L}, \mathit{cata}(X,T,Y1), {R}\}$;
 
\end{minipage}

\smallskip
\noindent
Then, by \textit{unfolding} $D$ we derive $S_{i+1}= (S_i \setminus \{D\}) \cup \textit{UnfCls}$. 

\smallskip

\noindent
For instance, in our Insertion Sort example, by unfolding definition $D1$, at Step~1 we replace $D1$ by:


\noindent
$E1$.~~$\mathit{new}1(A,B,A,C,[~]) \leftarrow \ordered(A,B),\ \ordered(A,C).$

\noindent
$E2$.~~$\mathit{new}1(A,B,C,D,[E|F]) \leftarrow \ordered(A,B),\ \ordered(C,D),$
          
\hspace*{44.5mm}$ \emptylist(G),\ \ordins(E,G,A,F,C).$

\noindent
Step 2 of the unfolding rule is not performed in this example. At Step 3, clause $E1$ is 
replaced by:


\noindent
$E3$.~~$\mathit{new}1(A,B,A,C,[~]) \leftarrow B\!=\!C, \ordered(A,B).$

\noindent
Thus, $S_{2}=S_0 \cup \{E2,E3\}$. \hfill$\Box$~~

\smallskip
The \textit{query-based strengthening rule} allows us to use the 
queries occurring in the set of CHCs 
whose satisfiability is under verification for strengthening the body
of the other clauses with the addition of catamorphism atoms 
and constraints.

\smallskip
\noindent
(R3)~{\it{Query-based Strengthening} Rule.}
Let $S_i = P \cup Q$, where $P$ is a set of definite clauses obtained by 
applying the unfolding rule, and $Q$ is a
set of catamorphism-based queries, and let $C$: $H\leftarrow c, \textit{Catas}^{{C}}\!, A_1,\ldots,A_m$ 
be a clause in $P$, being the~$A_{i}$'s program atoms. Let $E$ be the clause derived from $C$ as follows:


\noindent 
{\bf for} $k=1,\ldots,m$ {\bf do}

\vspace*{0mm}
\noindent
\begin{minipage}{121mm}
\noindent
\hangindent=3mm
\makebox[3mm]{-}consider program atom $A_{k}$; let $\mathit{Catas}_{k}^{{C}}$ be the conjunction of every catamorphism atom~$F$ 
 in $\textit{Catas}^{C}$ such that $\mathit{adt\mbox{-}vars}(A_k) \cap 
\mathit{adt\mbox{-}vars}(F) \neq \emptyset$;

\hangindent=3mm
\makebox[3mm]{-}\textbf{if} in $Q$ there exists a query (modulo variable renaming)
\\
$q_k$: $\textit{false} \leftarrow c_k, \mathit{cata}_1(X_1,T_1,Y_1),\ldots,
\mathit{cata}_n(X_n,T_n,Y_n), A_k$ 
\\
where $Y_1,\ldots,Y_n$ do not occur in~$C$, and the conjunction
$\mathit{cata}_1(X_1,T_1,Y_1),\ldots,$ 
$\mathit{cata}_n(X_n,T_n,Y_n)$ can be split into two
subconjunctions $B_1$ and $B_2$ such that:
(i)~a variant of $B_1$ is a subconjunction of $\mathit{Catas}^{C}_{k}$, and
(ii)~for every catamorphism atom $\mathit{cata}_i(X_i,T_i,Y_i)$ in $B_2$ 
there is no catamorphism atom 
$\mathit{cata}_i(V,T_i,W)$ in $\mathit{Catas}^{C}_k$
\\
\textbf{then} add the conjunction $\sim\!c_k, B_2$ to the body of $C$.

\end{minipage}

\smallskip

\noindent
Then, by \textit{query-based strengthening of clause}~$C$ \textit{using queries $q_{1},\ldots, q_{m}$} (some of these
queries may be absent), we get the new set $S_{i+1}= (S_i \setminus \{C\}) \cup \{E\}$. 

\smallskip

\noindent
For instance, from clause $E2$ by {query-based strengthening using $q_{2}$} (note that in~$E2$ the 
program atom $\emptylist$  has no associated query),  we get:


\noindent
$E4$.~~$new1(A,B,C,D,[E|F]) \leftarrow$\\
\indent
\hspace*{8mm}$\mathit{\big(L ~\ands~ B ~\ands~~ (( J \ands H ) 
\Rightarrow K\!\leq\! I) ~~\ands~~ ( J \Rightarrow K\!\leq\! E )\big) \ \Rightarrow \ D,}$\\
\indent
\hspace*{8mm}$\ordered(A,B),\ \ordered(C,D),$ \hfill$(B_{1})$\hspace*{8mm}\\
\indent
\hspace*{8mm}${\ordered(G,L)},\ {\last(G,J,K)},\ {\first(A,H,I)},$\hfill$(B_{2})$\hspace*{8mm}\\
\indent
\hspace*{8mm}$\emptylist(G),\ \ordins(E,G,A,F,C).$

\noindent
where the subconjunction $B_{1}$ mentioned in Rule R3 is in line $(B_{1})$, while
the subconjunction~$B_2$ 
is in line $(B_{2})$.
Thus, $S_{3}=S_0 \cup \{E3,E4\}$. \hfill$\Box$~~

\smallskip
The \textit{folding rule} allows us to replace a conjunction of
a program atom and catamorphisms by a single atom,
whose predicate has been introduced in a previous application of 
the Definition Rule. 

\smallskip
\noindent
(R4)~{\it Folding Rule.}
Let $C$: $H\leftarrow c, \textit{Catas}^{C}, A_1,\ldots,A_m$ 
be a clause in $S_i$, where either $H$ is \textit{false} or
$C$ has been obtained by the unfolding rule, possibly followed by 
query-based strengthening. 
For $k=1,\ldots,m,$

\smallskip
\noindent
\begin{minipage}{122mm}
\noindent
\hangindent=3mm
\makebox[3mm]{-}let $\mathit{Catas}_{k}^{C}$ be the conjunction of every catamorphism atom $F$ 
in $\textit{Catas}^{C}$ such that $\mathit{adt\mbox{-}vars}(A_k) \cap 
\mathit{adt\mbox{-}vars}(F) \neq \emptyset$;

\hangindent=3mm
\makebox[3mm]{-}let
$D_k$: $H_k \leftarrow d_k, \mathit{Catas}^{D}_{k}, A_k$ be a clause in $\mathit{Defs}_i$ (modulo variable renaming)
such that:
(i)~$\mathbb D\models \forall(c \rightarrow d_k)$, and (ii)~$\mathit{Catas}^{C}_{k}$ is a subconjunction of $\mathit{Catas}^D_{k}$.

\end{minipage}

\smallskip\noindent
Then, by \textit{folding \( C\)
	using~\( D_1,\ldots,D_m\)}, we derive clause 
\(E  \):~\( H\leftarrow c, H_1,\ldots,H_m \), and we get
\( S_{i+1}= (S_{i}\setminus\{C\})\cup \{E \} \).

\medskip

In order to fold clause $E4$, we introduce two new definitions, one for each
program atom occurring in the body of that clause, as follows (the predicate names are
introduced by our tool):

\noindent $D2$.~~$\mathit{new}2(A,B,C,D,E,F,G,H,I,J,K,L)\leftarrow 
\ordered(A,B),\ \ordered(C,D),$\\
 \hspace*{15mm} $\ordered(E,F),\  
          \last(A,G,H),\ \first(C,I,J),\ \ordins(K,A,C,L,E).$ \\[-4mm]
          
\noindent $D3$.~~$\mathit{new}19(A,B,C,D)\leftarrow 
\ordered(A,B),\ \last(A,C,D),\ \emptylist(A).$ 



Thus, $S_{4}\!=\!S_0 \cup \{E3,E4,D2,D3\}$. 

Now, we apply Rule~R4, and from clause~$E4$ we get:

\noindent $E5$.~~$\mathit{new}1(A,\!B,\!C,\!D,\![E|F]) \leftarrow 
\big(G\,\ands\, 
B\,\ands\, 
((H \ands I)\!\Rightarrow\! J \!\leq\! K) \,\ands\, 
(H\!\Rightarrow \! J \!\leq\! E) \big) \Rightarrow D, \\ 
  \hspace*{15mm}   \mathit{new}2(L,G,A,B,C,D,H,J,I,K,E,F), \ 
         \mathit{new}19(L,G,H,J).$ 


\noindent
Then, $S_{5}=S_0 \cup \{E3,E5,D2,D3\}$.
Also, query $q1$ can be folded using definition~$D1$, and we get:

\noindent $E6$.~~$\mathit{false\ \leftarrow\  \sim\!(B{\rm1}\Rightarrow B{\rm2}),\  new{\rm1}(Xs,B{\rm1},Zs,B{\rm2},Ys).}$ 


Thus, $S_{6}=(S_0 \setminus \{q1\}) \cup \{E3,E5,E6,D2,D3\}$.
Then, the transformation will continue by looking for the clauses 
relative to 
the newly introduced predicates $\mathit{new}2$ (see Definition~$D2$) and 
$\mathit{new}19$ (see Definition~$D3$).~\hfill$\Box$~~

\smallskip
The key for understanding our transformation technique is to
observe that in our Insertion Sort example, the query $E6$ and the clauses $E3$ and $E5$
that define the new predicate $\mathit{new}1$, incorporate the program
predicate $\inssort$ together with the
catamorphisms that are used in $q1$.
The advantage of performing this transformation is that the solver 
can look for a model of $\mathit{new{\rm1}(Xs,\!B{\rm1},\!Zs,\!B{\rm2},\!Ys)}$
where the constraint $B{\rm1}\!\!\Rightarrow\!\! B{\rm2}$ holds, instead of
looking in a separate way for models of 
$\ordered(Xs,\!B1)$,  $\ordered(Zs,\!B2)$,
and $\mathit{\inssort(Xs,\!Ys,\!Zs)}$ whose conjunction 
implies  $B{\rm1}\!\Rightarrow\! B{\rm2}$. Note also that clause~$E5$ has
constraints and catamorphisms that come from query-based strengthening.
Indeed,~$E5$ is obtained by folding $E4$ derived by strengthening $E2$  using query~$q2$.


The following Theorem~\ref{thm:Corr}, whose proof sketch is given in Appendix~1,
states the correctness of the 
transformation rules.

\vspace{-1mm}
\begin{theorem}[Soundness and Completeness of the Rules]
\label{thm:Corr}
Let $S_0 \TransfTo S_1 \TransfTo \ldots \TransfTo S_n$ be 
a transformation sequence
using rules {\rm{R1--R4}}.
Then, $S_0$ is satisfiable if and only if $S_n$ is satisfiable.
\end{theorem}

\vspace*{-2mm}

Note that the applicability conditions of R3 disallow
the application of the rule to a query. Otherwise, we
could easily get a satisfiable clause from an
unsatisfiable one. Indeed, we could transform\ 
$\mathit{false} \leftarrow c(Y), \mathit{cata}(X,Y), p(X)$
into
$\mathit{false} \leftarrow \ \sim\! c(Y)\,\ands\,c(Y), 
     \mathit{cata}(X,Y), p(X)$.
Note also that folding a clause using itself is not allowed, thus
avoiding the transformation of $H \leftarrow c, Catas, A$ into
the trivially satisfiable clause $H \leftarrow c, H$.

The applicability conditions of the rules force a sequence of
the transformation rules which is fixed, if the new definitions
to be introduced are known. The algorithm that we will present in the next section shows 
how these definitions can be introduced in an automatic way.

\section{Transformation Algorithm} 
\label{sec:Strategy}
In this section we present  
an algorithm, called~\mq, which given a set $P$ of
definite clauses and a set $Q$ of queries, introduces a 
set of 
new predicates and transforms $P\cup Q$  into a new set $P'\cup Q'$
such that: (i) $P\cup Q$ is satisfiable if and only if $P'\cup Q'$ is so,
and (ii) each new predicate defined in $P'\cup Q'$
is equivalent to the conjunction of a program predicate
and some catamorphisms needed for checking the satisfiability 
of the queries in $Q$. 
As an effect of the application of the query-based strengthening rule,
the transformed clauses also exploit the interdependencies among the 
queries in $Q$. This transformation is effective, in particular,
in the presence of mutually recursive predicates, like
in our Insertion Sort example, where we are able to get new clauses
for checking the satisfiability of the query $q1$ for $\inssort$ that take
into account
{the constraints and catamorphisms} 
of the query $q2$ for $\ordins$, and vice versa. 

The set of new definitions needed by~\mq~is computed
as the least fixpoint of an operator $\tau_{P,Q}$ that transforms a set $\Delta$ of
definitions into a new set $\Delta'$. For introducing that operator, we need some preliminary definitions
and functions.

\begin{definition}
A {\em generalization}  of a pair $(c_1,c_2)$ of constraints is a constraint, denoted $\alpha (c_1,c_2)$,
such that $\mathbb D \models\forall (c_1 \!\rightarrow \alpha (c_1,c_2)\!)$ and 
$\mathbb D \models\forall (c_2 \!\rightarrow \alpha (c_1,c_2)\!)$~$\cite{Fi&13a}$.
The {\em projection} of a constraint
$c$ onto a tuple $V$ of variables is a constraint $\pi(c,V)$ such that: (i)~$\mathit{vars}(\pi(c,V))\!\subseteq\! V$ and
(ii)~$\mathbb D \models \forall (c\!\rightarrow\!\pi(c,V))$. 
\end{definition} 

A set $\Delta$ of definitions is
\emph{monovariant} if it contains at most one definition for each program predicate.

\begin{definition}
Let $D_1$: $\mathit{newp}1(U_1) \If c_1, \mathit{Catas}_1, p(Z)$ 
and $D_2$: $\mathit{newp}2(U_2) \If c_2, \mathit{Catas}_2, p(Z)$ be two definitions for the same predicate $p$.
We say that $D_2$ is an \emph{extension} of $D_1$, written $D_1\sqsubseteq D_2$,
if 
$(i)$~$\mathit{Catas}_1$ is a subconjunction of $\mathit{Catas}_2$,  and
$(ii)$~\mbox{$\mathbb D\!\models\!\forall (c_1\! \rightarrow\! c_2)$}.
Let $\Delta_1$ and $\Delta_2$ be two monovariant sets of definitions.  
We say that 
$\Delta_2$ is an \emph{extension} of $\Delta_1$, written $\Delta_1\sqsubseteq \Delta_2$, if 
for each $D_1$ in $\Delta_1$ there exists~$D_2$ in $\Delta_2$ such that $D_1\sqsubseteq D_2$.

\end{definition}

\vspace{-1mm}
Given a set \textit{Cls} of clauses and a set $\Delta$ of definitions,
the {\it Define} function (see Figure~\ref{fig:Functions}) 
derives a set $\Delta'$ of definitions that can be 
used for folding all clauses in \textit{Cls}.
If $\Delta$ is monovariant,
then also $\Delta'$ is monovariant.
In particular, due to the (Project) case, $\Delta'$ contains a definition
for each program predicate occurring in the body of clauses in \textit{Cls}.
Due to the (Extend) case, $\Delta \sqsubseteq \Delta'$.

\begin{figure}[!ht]
\begin{minipage}{122mm}
\noindent \hrulefill \nopagebreak


\noindent {\bf Function $\mathit{Define}(Cls,\Delta)$}:
a set $\mathit{Cls}$ of clauses; a monovariant set $\Delta$ of definitions.
$\mathit{Define}(Cls,\Delta)$ returns a monovariant set $\Delta'$ of new definitions computed as follows.

\vspace{1mm}

\noindent $\Delta' := \Delta$;

\noindent {\bf for} each clause $C$: $H\leftarrow c, G$ in $\mathit{Cls}$ {\bf do}

\noindent 
\hspace{3mm}{\bf for} each program atom $A$ in $G$ {\bf do}

\smallskip

\hspace{6mm}
\begin{minipage}{2mm}
\end{minipage}
\begin{minipage}{110mm}
\noindent
let $\mathit{Catas}\!_A$ be the conjunction of every catamorphism atom $F$ in $G$ such that $\mathit{adt\mbox{-}vars}(A) \cap 
\mathit{adt\mbox{-}vars}(F) \neq \emptyset$

\smallskip

$\bullet$ ({\bf Skip}) {\bf if} in $\Delta'$ there is a clause $\mathit{newp}(U) \If d, 
    B, A$, for any conjunction $B$ of catamorphism atoms,
such that:
(i)~$\mathit{Catas}\!_A$ is a subconjunction of $B$,
and (ii)~$\mathbb D \models\forall (c \rightarrow d)$, 
{\bf then} \underline{\Down skip};

\smallskip

$\bullet$ ({\bf Extend}) {\bf else if} the definition for the predicate of $A$  in $\Delta'$
is the clause $D$: $\mathit{newp}(U) \If d, B, A$, where $B$ is a conjunction of catamorphism
atoms, and either
(i)~$\mathit{Catas}\!_A$ is {\em not} a subconjunction of $B$, or
(ii)~$\mathbb D \not\models\forall (c \rightarrow d)$,
{\bf then}

\underline{\Down introduce definition} $\mathit{ExtD}$: $\mathit{extp}(V) \If \alpha(d,c), A, B'$,
where: (i) $\mathit{extp}$ is a new predicate symbol,
(ii) $V\!=\!\mathit{vars}(\{\alpha(d,c),B',A\})$, and
$B'$ is the conjunction of the distinct catamorphism atoms occurring either in $B$ or in $\mathit{Catas}\!_A$; 

$\Delta' := (\Delta' \setminus \{D\}) \cup \{\mathit{ExtD}\}$; 

\smallskip

$\bullet$ ({\bf Project}) {\bf else if} there is no clause in $\Delta'$ of the form
$K \If d, B, A$, for any conjunction $B$ of catamorphism atoms,

{\bf then} \underline{\Down introduce definition} $D$: $\mathit{newp}(U) \If \pi(c,I), A, \mathit{Catas}\!_A$,
where: 
(i)~$\mathit{newp}$ is a new predicate symbol,
(ii)~$I$ are the input variables of basic sort in $\{A, \mathit{Catas}\!_A\}$, and
(iii)~$U\!=\!\mathit{vars}(\{\pi(c,I), A, \mathit{Catas}\!_A\})$;


$\Delta' := \Delta' \cup \{D\}$;

\smallskip

\end{minipage}

\noindent \hrulefill

\noindent {\bf Function $\mathit{Unfold}(\Delta,P)$}:\,a set $\Delta\!=\!\{D_1,\!\ldots\!,D_n\}$\,of definitions;\,a set\,$P$ of definite clauses.\\
$\mathit{Unfold}(\Delta,P)= \bigcup^n_{i=1} C_i$, where $C_i$ is the set of clauses derived by unfolding $D_i$.

\vspace*{-2mm} \noindent \hrulefill

\noindent {\bf Function $\mathit{Strengthen}(Cls,Q)$}: a set $\mathit{Cls}=\{C_1,\ldots,C_n\}$ of clauses;
a set $Q$ of catamor\-phism-based queries, at most one query for each program predicate {in $\mathit{Cls}$}. \\ 
$\mathit{Strengthen}(Cls,Q)\!=\!\{E_i\!\mid\!E_i$ is derived from $C_i$\,by query-based strengthening using $Q\}$

\vspace*{-2mm} \noindent \hrulefill

\noindent {\bf Function $\mathit{Fold}(Cls,\Delta)$}:
a set $\mathit{Cls}=\{C_1,\ldots,C_n\}$ of clauses; a monovariant set $\Delta$ of definitions.\\
$\mathit{Fold}(\mathit{Cls}, \Delta) = \{E_i \mid E_i$ is derived from $C_i$ by folding $C_i$ 
using definitions in $\Delta\}$.

\vspace*{-2mm} \noindent \hrulefill

\vspace*{-1mm}
\caption{The $\mathit{Define}$, $\mathit{Unfold}$, $\mathit{Strengthen}$, and $\mathit{Fold}$ 
functions. \label{fig:Functions}}
\end{minipage}
\vspace*{-6mm}
\end{figure}
The  $\mathit{Unfold}$ and $\mathit{Strengthen}$ functions  (see Figure~\ref{fig:Functions})
apply the unfolding and query-based strengthening
rules, respectively, to sets of clauses.

Now, we define the operator $\tau_{P,Q}$ as follows:


\vspace{.5mm}
$\tau_{P,Q}(\Delta) = \Bigg\{
\begin{matrix}
\mathit{Define}(Q,\emptyset) \hspace*{47mm} & \mbox{if}~\Delta=\emptyset  \\[1.mm]
\mathit{Define}(\mathit{Strengthen}(\mathit{Unfold}(\Delta,P),Q),\Delta)~~~~~ & ~\mbox{otherwise}
\end{matrix}$

\vspace{.5mm}
\noindent
In the case where $\Delta$ is the empty set of definitions, $\tau_{P,Q}(\Delta)$ 
introduces by the \textit{Define} function (Project case),  a new definition for each program
predicate occurring in a query in $Q$.
In the case where $\Delta$ is not empty, 
$\tau_{P,Q}(\Delta)$ is an extension of $\Delta$ obtained by 
first unfolding all clauses in $\Delta$, 
then applying the query-based strengthening rule to the clauses derived by unfolding,
and finally applying the \textit{Define} function. 

The \textit{Define} function is parametric with respect to the 
generalization operator~$\alpha$ (see the Extend case).
In our implementation we use an operator based on widening~\cite{Fi&13a}
that ensures \textit{stabilization}, that is,
for any infinite sequence $c_0,c_1,\ldots$ of constraints and any sequence defined as
(i)~$d_0\!=\!c_0$, and (ii)~$d_{k+1}\!=\!\alpha(d_k,c_{k+1})$,
there exists $m\!\geq\! 0$ such that $d_m\!=\!d_{m+1}$.
The \textit{Strengthen} function ensures that there is a bound on the number
of catamorphisms that can be present in a definition.
Since, as already mentioned, $\tau_{P,Q}$ is monotonic with respect to $\sqsubseteq$,
its least fixpoint $\textit{lfp}(\tau_{P,Q})$ 
is equal to $\tau_{P,Q}^n(\emptyset)$, for some finite number $n$ of iterations of 
$\tau_{P,Q}$.
Note that, by construction, $\textit{lfp}(\tau_{P,Q})$ is monovariant.

Once we have computed the set $\textit{lfp}(\tau_{P,Q})$ of definitions, 
we can use them to fold all clauses derived by unfolding and strengthening
by applying the \textit{Fold} function (see Figure~\ref{fig:Functions}).
Thus, the transformation algorithm is defined as follows:

\vspace{1mm}
\mq$(P,Q) = \mathit{Fold}\big(\mathit{Strengthen}\big(\mathit{Unfold}(\mathit{lfp}(\tau_{P,Q}),P),\,Q\big),\ \mathit{lfp}(\tau_{P,Q})\big)$



\vspace{1mm}
\noindent
Termination of \mq~follows immediately from
the fact that $\textit{lfp}(\tau_{P,Q})$ is computed in a finite number of steps.

\vspace{-2mm}
\begin{theorem}[Termination of {Algorithm}~\mq]  
\label{thm:termination}
Let $\mathit{P}$ be a set of definite clauses and $Q$ a set of catamorphism-based
queries.
Then, Algorithm~\mq~terminates for $\mathit{P}$ and $Q$.
\end{theorem}

\vspace{-2mm}
By the soundness and completeness of
the transformation rules (see Theorem~\ref{thm:Corr}), we also
get the following result.

\vspace{-2mm}
\begin{theorem}[Soundness and Completeness of {Algorithm}~\mq] 
\label{thm:soundness-AlgorithmR}
For any set $\mathit{P}$ of definite clauses and $Q$ of catamorphism-based queries,
 $P \cup Q$ is satisfiable if and only if
\mq$(P,Q)$ is satisfiable.
\end{theorem}

\vspace{-2mm}

We conclude this section by showing the sequence of definitions for the program predicate
\textit{snoc} computed by iterating the applications of $\tau_{P,Q}$ 
in the Insertion Sort example. The definitions in $\textit{lfp}(\tau_{P,Q})$ are listed in 
Appendix~2.

\vspace*{1mm}
{\small{
\noindent\hspace*{5mm}$\mathit{new}4(A,B,C,D,E,F,G)  \leftarrow \ordered(A,B),\ \last(A,C,D),\ 
      \ordered(E,F), \\[-.5mm]
     \hspace*{15mm} \snoc(A,G,E).$\\[-.5mm] 
\noindent\hspace*{5mm}$\mathit{new}5(A,B,C,D,E,F,G,H,I,J,K)  \leftarrow 
       \ordered(E,J),\ \last(E,F,G), \\[-.5mm]
      \hspace*{15mm} \ordered(A,D),\ \first(E,H,I),\ \first(A,B,C),\ \snoc(E,K,A).$\\[-.5mm] 
\noindent\hspace*{5mm}$\mathit{new}13(A,B,C,D,E,F,G,H,I,J,K,L,M) \leftarrow \ordered(A,B),\ 
        \last(A,C,D), \\[-.5mm] 
      \hspace*{15mm}  \ordered(E,F),\ \last(E,G,H),\  \first(E,J,K),\ \first(A,L,M),\ \snoc(A,I,E).$
}}

\vspace*{1mm}
Note that these three definitions are in the $\sqsubseteq$ relation.
\vspace*{-2mm}

\section{Experimental Evaluation} 
\label{sec:Experiments}

We have implemented algorithm \mq~in a tool, called \vericatmq, which extends \vericat~\cite{DeAngelisFPP22}
by guaranteeing a sound and complete transformation.
\vericatmq~is based on (i) VeriMAP~\cite{De&14b} for transforming CHCs, 
and (ii) SPACER (with Z3 4.11.2) to check the satisfiability 
of the transformed CHCs.



We have considered 170 problems, as sets of CHCs, with 470 queries in total,
equally divided between the class of satisfiable (sat) problems and unsatisfiable (unsat) ones
(85~problems and 235~queries for each class).
These problems are related to programs
that manipulate: (i)~lists of integers by performing concatenation, permutation, reversal and sorting,
and (ii)~binary search trees, by inserting and deleting elements.
%
For list manipulating programs, we have considered properties such as: list length, minimum and maximum element, sum of elements, list content as sets or
 multisets of elements, and list sortedness (in ascending or descending order).
For trees, we have considered size, height, minimum and maximum element, tree content and the 
 {binary search tree property}.

The problems considered here are derived from those of the benchmark set of previous work~\cite{DeAngelisFPP22} 
with some important differences.
We have considered additional satisfiable problems (for instance, those related to Heapsort).
%
In addition to satisfiable problems, we have also considered unsatisfiable problems
that have been obtained from their satisfiable counterparts
by introducing bugs in the programs: 
for instance, by not inserting an element in a list, or adding an extra constraint,  
or replacing a non-empty tree by an empty one.
%
Note also that the transformed CHCs produced by  \mq~contain both basic variables and ADT variables, whereas 
those produced by the method presented in previous work~\cite{DeAngelisFPP22} 
contain basic variables only.

For comparing the effectiveness of our method with that of a state-of-the-art CHC solver,
we have also run  SPACER (with Z3 4.11.2)
on the original, non-transformed CHCs, in SMT-LIB format.
In Table~\ref{tab:exp} we summarize the results of our 
experiments\footnote{Experiments have been performed on an Intel Xeon CPU E5-2640 2.00GHz with 64GB RAM under CentOS with a time limit of 300s per problem.}\!.
The first three columns report the name of the program, the total number of problems
and queries
for each program. 
The fourth and fifth columns report the number of satisfiable and unsatisfiable problems 
proved by SPACER before transformation,
whereas the last two columns report the number of satisfiable and unsatisfiable problems 
proved by \vericatmq.

{
In summary,
\vericatmq was able to prove all the 170 considered problems 
whereas SPACER was able to prove the properties of 84 `unsat' problems out of 85, and none of the `sat' problems.
The total time needed for transforming the CHCs was 275 seconds (1.62\,s per problem, on average),
and 
checking the satisfiability of the transformed CHCs took about 174\,s in total (about 1s average time,  0.10\,s median time).
%
For comparison, SPACER took 30.27\,s for checking the unsatisfiability of 84 problems  (0.36\,s average time,  0.15\,s median time). 
The benchmark and the tool are available at \url{https://fmlab.unich.it/vericatmq}.
}

\begin{table}
\centering
\begin{tabular}{|@{\hspace{1mm}} l ||r@{\hspace{2mm}}|r@{\hspace{2mm}}|r@{\hspace{2mm}}|r@{\hspace{2mm}}|r@{\hspace{2mm}}|r@{\hspace{2mm}}|}
\hline
{\parbox[center]{35mm}{Program}}       & 
{\parbox[top]{12mm}{Problems}} &  {\parbox[top]{10mm}{Queries}} & 
\multicolumn{2}{c|}{SPACER}  &
\multicolumn{2}{c|}{\vericatmq}  \\
        & 
 &   & 
~\textrm{sat} & ~\textrm{unsat} &
~\textrm{sat} & ~\textrm{unsat}  \\
\hline\hline
List Membership \rule{0mm}{3.5mm} & 2 & 6              &  0   & 1   &  1  & 1  \\[-.5mm]
List Permutation      & 8 & 24                         &   0  &  4   & 4   & 4  \\[-.5mm]
List Concatenation   & 18 &18                         &   0  & 8   &   9  & 9  \\[-.5mm]
Reverse               & 20 & 40                         &   0  & 10   & 10   &  10 \\[-.5mm]
Double Reverse        & 4 & 12                          &   0  & 2   &  2  & 2  \\[-.5mm]
Reverse w/Accumulator & 6 & 18                 &  0   & 3   & 3   & 3  \\[-.5mm]
Bubblesort           & 12 & 36                         &  0   &  6  &  6  & 6  \\[-.5mm]
Heapsort        & 8 & 48                        	  &   0  &  4  &  4   & 4  \\[-.5mm]
Insertionsort        & 12 & 24                         &  0   &   6 & 6   &  6 \\[-.5mm]
Mergesort            & 18 & 84                         &  0   &  9  & 9   &  9 \\[-.5mm]
Quicksort (version 1)   & 12 & 38                      &  0   &   6 & 6   &  6 \\[-.5mm]
Quicksort (version 2)   & 12 & 36                      &  0   &   6 & 6   &  6 \\[-.5mm]
Selectionsort        & 14 & 42                        &  0   &   7 & 7   &  7 \\[-.5mm]
Treesort             & 4 & 20                         &   0  & 2   &  2  & 2  \\[-.5mm]
Binary Search Tree    & 20 & 24                         &   0  & 10   &  10  & 10  \\[-.5mm]
\hline\hline
\textrm{Total} \rule{0mm}{3.mm}&	\textrm{170} & \textrm{470} & 0 & 84 & 85 &  85 \\
\hline
\end{tabular}
\vspace{2mm}
\caption{Programs and problems proved by SPACER and \vericatmq.}
\label{tab:exp}
\vspace*{-9mm}
\end{table}

For instance, for all the considered list sorting programs 
(Bubblesort, Heapsort, Insertionsort, Mergesort, Quicksort, Selectionsort and Treesort),
\vericatmq was able to prove properties stating that the output list is sorted  
and has the same multiset of elements of the input list.
Similarly, \vericatmq was able to prove that those properties do \textit{not} hold,
if extra elements are added to the output list, or some elements are not copied from the input list to the output list, 
or a wrong comparison operator is used.

The results obtained by the \vericatmq prototype implementation of our method are encouraging
and show that, when used in combination with state-of-the-art CHC solvers,
it can greatly improve their effectiveness to prove satisfiability 
of sets of CHCs with multiple queries,  
while it does not inhibit their remarkable ability to prove unsatisfiability, {although some extra time due to transformation may be required}.

\section{Conclusions and Related Work} 
\label{sec:RelConcl}
\vspace*{-2mm}
Many program verification problems can be
translated into the satisfiability problem for sets of CHCs that
include more than one query. A notable example is the case where we want to verify
the correctness of programs made out of several functions,
each of which has its pre-/postconditions~\cite{Gr&12}.
We have proposed an algorithm, called \mq,
for transforming a set of CHCs with multiple queries
into a new, equisatisfiable set of CHCs
{that incorporate suitable information
about the set of queries contained in the initial set.}
The advantage gained is that, in order to prove the satisfiability of the transformed CHCs,
the CHC solver may exploit the mutual interactions among the satisfiability proofs
of the various queries. We have identified
a class of queries that specify program properties using catamorphisms
on ADTs, such as lists and trees, for which \mq~terminates.
We have implemented algorithm~\mq~and 
shown that it improves the effectiveness of the state-of-the-art
CHC solver SPACER~\cite{Ko&16} on a non trivial benchmark.

Algorithm \mq~improves over the transformation algorithm~\Cata~presented in a
previous paper~\cite{DeAngelisFPP22}, which works by eliminating ADTs from sets of CHCs.
Instead of the contracts handled by \Cata, algorithm \mq~considers queries, 
and thus its input is simply a set of CHCs.
More importantly, \mq~is sound and complete, in the sense that
the initial and transformed CHCs are equisatisfiable sets, whereas 
\Cata~is only sound (that is, it can be seen as computing an \textit{abstraction}
of the initial clauses), and thus if the transformed clauses are unsatisfiable
we cannot infer anything about the satisfiability of the initial CHCs.
Completeness is very important in practice, because proving
that a set of clauses is unsatisfiable and finding a counterexample
can help identify a program bug.
The experimental evaluation reported in Section~\ref{sec:Experiments}
shows that our transformation-based verification technique
is able to dramatically improve the effectiveness of the SPACER solver
for satisfiable sets of CHCs (where the results of SPACER are very poor), 
while retaining the excellent results of the solver for unsatisfiable
sets of CHCs (for which SPACER is, at least in principle, complete).

Decision procedures for suitable classes of first order formulas
defined on catamorphisms~\cite{PhamGW16,SuterDK10} have been used in
program verifiers~\cite{Su&11}.
However, we do not propose here any specific decision procedure for
catamorphisms and, instead, we transform a set of CHCs with catamorphisms 
into a new set of CHCs where catamorphisms are, in a sense, compiled away.

{\em Type-based norms}, which are a special kind of integer-valued catamorphisms, 
were used for proving termination of logic programs~\cite{BruynoogheCGGV07} and for
{\em resource analysis}~\cite{AlbertGGM20} via \textit{abstract interpretation}.
Similar abstract interpretation techniques are also implemented in
the CiaoPP preprocessor~\cite{HermenegildoPBL05} of the Ciao logic programming system.
In our approach we do not need to specify \textit{a priory} any abstract domain where to perform
the analysis, and instead, by transformation, we generate new CHCs which incorporate
the relations defined by the constraints on the catamorphisms.
The problem of showing the satisfiability of CHCs defined on ADTs is a very hot topic
and various approaches have been proposed in recent work, including:
(i)~a proof system that combines inductive theorem proving with
CHC solving~\cite{Un&17},
(ii)~lemma generation based on syntax-guided synthesis from user-specified templates~\cite{Ya&19},
(iii)~invariant discovery based on finite tree automata~\cite{KostyukovMF21},
and (iv)~use of suitable abstractions~\cite{GovindSG22}.

A limitation of our approach is that the effectiveness of the transformation
may depend on the set of properties specified through the queries.
For instance, it may happen that programmers provide \textit{partial}
program specifications (e.g., for a subset of the program functions),
and therefore queries only for some program predicates, such as the 
main program predicates (e.g., $\mathit{\inssort}$ and $\mathit{\ordins}$ 
of our example).
In this case, it is essential to have a mechanism that is able to infer
from the queries the unspecified catamorphisms for the remaining program
predicates (e.g., $\mathit{append}$ and $\mathit{snoc}$ which 
$\mathit{\inssort}$ and $\mathit{\ordins}$ depend on).
As future work, we plan to extend \mq~to propagate the catamorphisms 
specified in the queries
to those program predicates for which no query has been specified.


    
\appendix

\newpage
\section*{Appendix 1}
\label{app:RulesTheorem}
\noindent 
In this appendix we present a proof sketch of Theorem~\ref{thm:Corr} 
that states the soundness and completeness of the transformation rules.

\smallskip
\noindent
{\it Proof sketch}. 
The proof of Theorem~\ref{thm:Corr} is based on the correctness of
the transformation rules for (constraint) logic programs~\cite{EtG96,TaS86}.
In particular, the addition of catamorphisms performed by the
query-based strengthening rule, is sound and complete
because the catamorphisms are total, functional relations. 
The correctness of
folding is proved by using a method similar to the
one introduced by Tamaki and Sato~\cite{TaS86}, which 
relies on two facts: (i) we can associate predicates with
\textit{levels} where program predicates have a higher level
than catamorphisms, and (ii) only clauses obtained by unfolding
with respect to a program atom are folded.

\section*{Appendix 2}
\label{app:Defs-FinalProg}
\noindent 
In this appendix let us first show the seven definitions that have been introduced during the transformation of the given set of CHCs presented in Figures~\ref{fig:chc_prog}, 
\ref{fig:chc_catas}, and~\ref{fig:chc_queries}. Note that in our case, the constraints in these definitions are all ${\mathit{true}}$.
 
\smallskip

\noindent
$D1.$~$\mathit{\mathit{new}}1(A,B,C,D,E) \leftarrow \ordered(A,B),\ \ordered(C,D),\ \inssort(A,E,C).$\\[-0pt]
$D2.$~$\mathit{new}2(A,B,C,D,E,F,G,H,I,J,K,L) \leftarrow \ordered(A,B),\ \ordered(C,D), 
\\ \hspace*{15mm}\ordered(E,F),\ \last(A,G,H),\ \first(C,I,J),\ \ordins(K,A,C,L,E).$\\[-0pt]
\hspace*{7mm}$\mathit{new}6(A,B,C,D) \leftarrow \ordered(A,B),\ \first(A,C,D).$\\[-0pt]
\hspace*{7mm}$\mathit{new}7(A,B,C,D,E,F,G,H,I,J,K,L,M,N) \leftarrow \first(A,B,C),\\ 
 \hspace*{15mm} \ordered(A,D),\ \last(E,F,G),\ \first(H,I,J),\ \ordered(H,K),  \\
\hspace*{15mm}  \first(E,L,M),\ \ordered(E,N),\ \append(E,A,H).$\\[-0pt] 
\hspace*{7mm}$\mathit{new}13(A,B,C,D,E,F,G,H,I,J,K,L,M) \leftarrow \ordered(A,B),  \\ 
 \hspace*{15mm}  \last(A,C,D),\ \ordered(E,F),\ \last(E,G,H),\ \first(E,J,K), \\
 \hspace*{15mm}  \first(A,L,M),\ \snoc(A,I,E).$\\[-0pt]
\hspace*{7mm}$\mathit{new}17(A,B) \leftarrow \ordered(A,B).$\\[-0pt]
$D3.$~$\mathit{new}19(A,B,C,D) \leftarrow \ordered(A,B),\ \last(A,C,D),\ \emptylist(A).$

\smallskip

Now we list the final set of CHCs which has been derived by our transformation technique using the 
above definitions. The first four CHCs are the queries derived from the queries $q1$--$q4$ of Figure~\ref{fig:chc_queries} denoting the four properties for Insertion Sort 
we wanted to prove.
We have that this final set of clauses is sastisfiable and thus, the four properties are all valid.

\smallskip
\noindent
$E6.~\mathit{false} \leftarrow \ \sim\!(A\Rightarrow B), \mathit{new}1(C,A,D,B,E).$\\
\hspace*{7mm}$\mathit{false} \leftarrow \  \sim\! ((A\,\ands\, B\,\ands\, ((C\,\ands\, D)\Rightarrow E\!\leq\! F)\,\ands\, (C\Rightarrow E\!\leq\! G))\Rightarrow H), \\
   \hspace*{14mm}  \mathit{new}2(I,A,J,B,K,H,C,E,D,F,G,L).$\\       
\hspace*{7mm}$\mathit{false} \leftarrow\  \sim\! (A\,\ands\, B\,\ands\, ((C\,\ands\, D)\Rightarrow E\!\leq\! F)
   \Rightarrow G),\\
   \hspace*{14mm}  \mathit{new}7(H,D,F,B,I,C,E,J,K,L,G,M,N,A).$\\
\hspace*{7mm}$\mathit{false} \leftarrow\ \sim\! ((A\,\ands\, (B\Rightarrow C\!\leq\! D))\Rightarrow E), \mathit{new}13(F,\!A,\!B,\!C,\!G,\!E,\!H,\!I,\!D,\!J,\!K,\!L,\!M).$\\
\hspace*{7mm}$\mathit{new}1(A,B,A,B,[~])\  \leftarrow\ \mathit{new}17(A,B).$\\
$E5.~\mathit{new}1(A,B,C,D,[E|F])  \leftarrow (G \ands B\,\ands\, ((H \ands I)\!\Rightarrow\! J\!\leq\! K)\,\ands\, (H\!\Rightarrow\! J\!\leq\! E))\Rightarrow D, \nopagebreak \\
     \hspace*{14mm}     \mathit{new}2(L,G,A,B,C,D,H,J,I,K,E,F),\ \mathit{new}19(L,G,H,J).$\\
\hspace*{7mm}$\mathit{new}2(A,B,[~],C,D,E,F,G,H,I,J,K)\  \leftarrow\ C \,\ands\, \sim\! H\,\ands\, I\!=\!0 \,\ands\, \\
       \hspace*{14mm}  ((B\,\ands\, (F\Rightarrow G\!\leq\! J))\Rightarrow L) \,\ands\, (L\Rightarrow E),\   
       \mathit{new}1(M,L,D,E,K),\\
     \hspace*{14mm}     \mathit{new}13(A,B,F,G,M,L,N,O,J,P,Q,R,S).$\\      
\hspace*{7mm}$\mathit{new}2(A,B,[C|D],E,F,G,H,I,J,K,L,M)\  \leftarrow\ (L\!\leq\! C) \,\ands\, \\
     \hspace*{14mm}   (E\!=\!(N\Rightarrow (C\!\leq\! O\,\ands\, P))) \,\ands\, J\,\ands\, K\!=\!C \,\ands\,
      ((B\,\ands\, (H\Rightarrow I\!\leq\! L))\Rightarrow Q) \,\ands\,\\  \hspace*{14mm}  ((Q\,\ands\, (R\Rightarrow S\!\leq\! C))\Rightarrow T)\,\ands\, 
      ((T\,\ands\, P\,\ands\, ((U\,\ands\, N)\Rightarrow V\!\leq\! O)) \Rightarrow W) \,\ands\,\\
    \hspace*{14mm} (W\Rightarrow G),\  \mathit{new}1(X,W,F,G,M),\\ 
      \hspace*{14mm} \mathit{new}7(D,N,O,P,Y,U,V,X,Z,A1,W,B1,C1,T), \\
     \hspace*{14mm}     \mathit{new}13(D1,Q,R,S,Y,T,U,V,C,E1,F1,G1,H1), \\
     \hspace*{14mm}    \mathit{new}13(A,B,H,I,D1,Q,R,S,L,I1,J1,K1,L1).$ \\     
\noindent
\hspace*{7mm}$\mathit{new}2(A,B,[C|D],E,F,G,H,I,J,K,L,M)\  \leftarrow\ (L\geq C\!+\!1)\,\ands\, \\
\hspace*{14mm} 
 (E\!=\!(N\Rightarrow (C\!\leq\! O\,\ands\, P))) \,\ands\,  J\,\ands\, K\!=\!C \,\ands\,  
         ((B\,\ands\, (H\Rightarrow I\!\leq\! C))\Rightarrow Q) \,\ands\, \\
   \hspace*{14mm}   ((Q\,\ands\, P\,\ands\, (((R\,\ands\, N)\Rightarrow S\!\leq\! O)\,\ands\, (R\Rightarrow S\!\leq\! L)))\Rightarrow G), \\ \nopagebreak
     \hspace*{14mm}     \mathit{new}2(T,Q,D,P,F,G,R,S,N,O,L,M),\\
     \hspace*{14mm} \mathit{new}13(A,B,H,I,T,Q,R,S,C,U,V,W,X).$\\
\hspace*{7mm}$\mathit{new}6([~],A,B,C)\  \leftarrow\ A \,\ands\, \sim \! B\,\ands\,C\!=\!0.$\\
\hspace*{7mm}$\mathit{new}6([A|B],\!C,\!D,\!E)  \leftarrow C\!=\!(F\!\Rightarrow\! (A\!\leq\! G\,\ands\,H)) \,\ands\,D\,\ands\,E\!=\!A,  \mathit{new}6(B,\!H,\!F,\!G).$
 \noindent           
\hspace*{7mm}$\mathit{new}7(A,\!B,\!C,\!D,\![~],\!E,\!F,\!A,\!B,\!C,\!D,\!G,\!H,\!I)\  \leftarrow\ \sim\! E\,\ands\, F\!=\!0\,\ands\, \sim\! G\,\ands\, H\!=\!0\,\ands\,I, \\
  \hspace*{12mm}  \mathit{new}6(A,D,B,C).$\\
 \noindent     
\hspace*{7mm}$\mathit{new}7(A,\!B,\!C,\!D,\![E|F],\!G,\!H,\![E|I],\!J,\!K,\!L,\!M,\!N,\!O)\  \leftarrow\ G\,\ands\, H\!=\!ite(P,Q,E)\, \ands\, \\
  \hspace*{12mm} 
     J\,\ands\, K\!=\!E\,\ands\, L\!=\!(R\Rightarrow (E\!\leq\! S\,\ands\, T))\, 
     \ands\, M\,\ands\, N\!=\!E\,\ands\,\\
  \hspace*{12mm}     O\!=\!(U\Rightarrow (E\!\leq\! V\,\ands\, W))\, \ands\, 
          ((W\,\ands\, D\,\ands\, ((P\,\ands\, B)\Rightarrow Q\!\leq\! C))\Rightarrow T),\ \\
       \hspace*{12mm} \mathit{new}7(A,B,C,D,F,P,Q,I,R,S,T,U,V,W).$\\      
           \noindent    
\noindent
\hspace*{7mm}$\mathit{new}13([~],A,B,C,[D],E,F,G,D,H,I,J,K)\  \leftarrow\ A\,\ands\, \sim\! B\,\ands\, C\!=\!0 \,\ands\, \\
       \hspace*{12mm}  E\!=\!(L\Rightarrow (D\!\leq\! M\,\ands\, N)) \,\ands\, 
                   F\,\ands\, G\!=\!ite(O,P,D)\,\ands\, H\,\ands\, I\!=\!D\,\ands\,\\
       \hspace*{12mm} \sim\! J\,\ands\, K\!=\!0\,\ands\, N\,\ands\, \sim\! L\,\ands\, M\!=\!0\,\ands\, \sim\!O\,\ands\, P\!=\!0.$\\
 \noindent    
\hspace*{7mm}$\mathit{new}13([A|B],\!C,\!D,\!E,\![A|F],\!G,\!H,\!I,\!J,\!K,\!L,\!M,\!N)  \leftarrow 
C\!=\!(O\Rightarrow (A\!\leq\! P\,\ands\, Q)) \,\ands\, \\
      \hspace*{12mm}    D\,\ands\, E\!=\!ite(R,S,A)\,\ands\, G\!=\!(T\Rightarrow (A\!\leq\! U\,\ands\, V))
      \,\ands\, H\,\ands\, I\!=\!ite(W,X,A)\,\ands\,\\
      \hspace*{12mm} K\,\ands\, L\!=\!A\,\ands\, 
          M\,\ands\, N\!=\!A\,\ands\, ((Q\,\ands\, (R\Rightarrow S\!\leq\! J))\Rightarrow V), \\
          \hspace*{12mm} \mathit{new}13(B,Q,R,S,F,V,W,X,J,T,U,O,P).$\\
\hspace*{7mm}$\mathit{new}17([~],A)\  \leftarrow\ A.$\\
\hspace*{7mm}$\mathit{new}17([A|B],C)\  \leftarrow\ C\!=\!(D\Rightarrow (A\!\leq\! E\,\ands\, F)), \mathit{new}6(B,F,D,E).$\\
\hspace*{7mm}$\mathit{new}19([~],A,B,C)\  \leftarrow\ A\,\ands\, \sim\! B\,\ands\, C\!=\!0.$\\

\noindent\hrule

\end{document}